\documentclass[twocolumn,prl,showpacs,floatfix]{revtex4}

\usepackage{graphicx}
\usepackage{dcolumn}
\usepackage{bm}
\usepackage{amsmath}

\begin{document}
 
\newcommand{\beq}{\begin{equation}}
\newcommand{\eeq}{\end{equation}}
\newcommand{\G}{\mbox{$\Gamma$}}
\newcommand{\e}{\mbox{$\epsilon$}}
\newcommand{\la}{\mbox{$\lambda$}}
\newcommand{\de}{\mbox{$\delta$}}
\newcommand{\w}{\mbox{$\omega$}}
\newcommand{\wo}{\mbox{$\omega_0$}}
\newcommand{\h}{\mbox{$\hbar$}}
 
\title{Fragmentation of the photoabsorption strength
in neutral and charged metal microclusters}
 

\author{C. Yannouleas} 
\altaffiliation{Current address: Joint Institute for Heavy Ion 
Research, Oak Ridge National Laboratory, Oak Ridge, Tennessee 37831.} 

\author{R.A. Broglia}
\affiliation{Dipartimento di Fisica, Universita di Milano, and
INFN, Sez. Milano, I-20133 Milano, Italy, and
The Niels Bohr Institute, DK-2100 Copenhagen \O, Denmark}

\author{M. Brack}
\altaffiliation{Permanent address: Institut f\"ur Theoretische Physik,
Universit\"at Regensburg, D-8400 Regensburg, West Germany.}
\affiliation{The Niels Bohr Institute, DK-2100 Copenhagen \O, Denmark}

\author{P.-F. Bortignon}
\affiliation{Istituto di Ingegneria Nucleare, CESNEF Politecnico 
di Milano, Italy, and INFN, LNL, Legnaro, Italy}

\date{16 December 1988; Phys. Rev. Lett. {\bf 63}, 255 (1989)} 

\begin{abstract}
The line shape of the plasma resonance in both
neutral and charged small sodium clusters is calculated. 
The overall properties of 
the multipeak structure observed in the photoabsorption cross section of
spherical Na$_8$ and Na$_{20}$ neutral clusters can be understood
in terms of Landau damping.
Quantal configurations are shown to play an important role.
In the case of charged Na$_9^+$ and Na$_{21}^+$ clusters a single
peak is predicted that carries most of the oscillator strength.
\end{abstract}

\pacs{36.40.+d, 31.50.+w, 33.20.Kf} 

\maketitle
 
Recent photoabsorption experiments \cite{heer} have revealed the surface
plasmon in small neutral sodium clusters. A qualitative
overall account of the dependence
of the resonance frequencies on the cluster size
can be achieved in terms of the extended ellipsoidal shell model
\cite{nils,clem}
and of the experimental static polarizabilities. The reported
line broadening seems to obtain, as suggested in ref. \cite{heer},
an important contribution
from the coupling of the dipole resonance to quadrupole shape
fluctuations of the cluster \cite{bt,pb}.
 
Due to the fact that no complete photoabsorption curve has yet
been experimentally determined, the detailed shapes and widths
of the resonances are still rather uncertain.
In particular, the contribution of direct plasmon decay mechanisms
to the line widths must be investigated, especially in the case of
small clusters, in view of the detailed microscopic
calculations carried out by Ekardt \cite{ekar} and Beck \cite{beck}
within the framework of the time-dependent local density
approximation (TDLA).
 
In the present paper, we study the structure of the
plasmon resonance within the mean-field framework making use of the
random phase approximation (RPA). Because these studies can most
clearly be done in the case of spherical clusters, we presently restrict
our analysis to the neutral Na$_8$ and Na$_{20}$ and to the charged
Na$_9^+$ and Na$_{21}^+$. It will be concluded that,
in the case of the neutral Na$_8$ and Na$_{20}$, 
Landau damping is important and that the observed line
shapes are the result of a detailed interplay between quantum size
effects and residual interactions among
particle-hole excitations, as well as of the thermal
fluctuations of the cluster shapes. In the case of the charged
Na$_9^+$ and Na$_{21}^+$, Landau damping is practically absent
and the dipole strength essentially consists of a single peak
exhausting most of the plasmon oscillator strength.
 
\begin{figure}[b]
\centering\includegraphics[width=8cm]{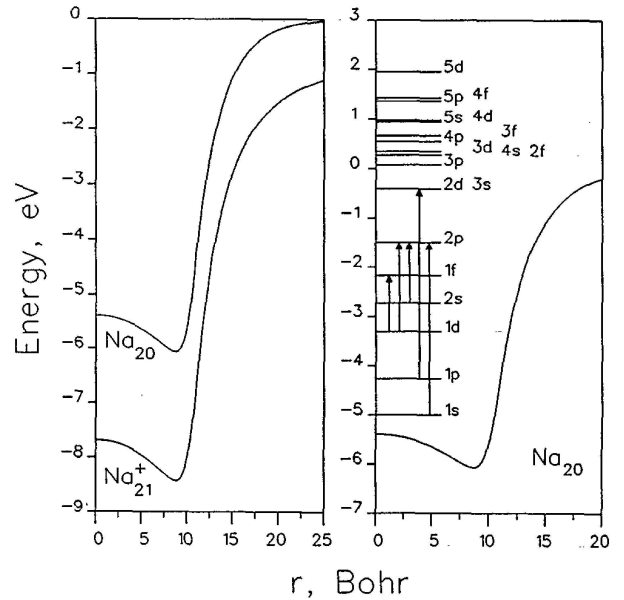}
\caption{
Self-consistent potentials \cite{brac} 
associated with Na$_{20}$ and Na$_{21}^+$ and used in the solution
of the single-particle
Hamiltonian $H_0$ appearing in eq. (\ref{eq:ham}).
For Na$_{20}$, the resulting single-particle levels 
relevant to the discussion of
the dipole resonance, as well as some of the associated unperturbed
particle-hole transitions, are also shown.}
\end{figure}
 
The calculations to be discussed below parallel those carried out in
studies of giant resonances in nuclei (cf., i.e. \cite{vang}). A discrete
particle-hole basis of dimension $\cal N$ is constructed and
the Hamiltonian
\beq
H=H_0+V,
\label{eq:ham}
\eeq
sum of the Hatree-Fock Hamiltonian $H_0$ and the
residual interaction $V$, is diagonalized using the RPA. In what
follows, we shall specify the single-particle potentials 
self-consistently in the spherical jellium-background
model \cite{eka2} using the density variational formalism in a
semiclassical approximation \cite{brac}. These potentials
are displayed in Fig.~1.

The residual two-body interaction $V$ is given by
\beq
V(|\vec{r}_1-\vec{r}_2|) =
\frac{e^2}{|\vec{r}_1-\vec{r}_2|}\;\;+
\;\;\frac{dV_{xc}[\rho]}{d\rho}\;\;\de (|\vec{r}_1-\vec{r}_2|)~.
\label{eq:rint}
\eeq
\noindent
Here $V_{xc}[\rho] = d{\cal E}_{xc}[\rho]/d\rho$ is the
exchange-correlation potential in the ground state. As in
refs. \cite{eka2,brac}, we use the exchange-correlation energy density
${\cal E}_{xc}[\rho]$ of Gunnarsson and Lundqvist \cite{gulu}.
 
The RPA equations (see, e.g., ref. \cite{rowe})
\begin{eqnarray*}
\left(
\begin{array}{cc}
A & B \\ B^* & A^*
\end{array}
\right)
\left(
\begin{array}{c}
X_n \\ Y_n
\end{array}
\right)
&  =  &
E_n
\left(
\begin{array}{c}
X_n \\ -Y_n
\end{array}
\right)~,  
\end{eqnarray*}
are written in terms of the angular-momentum coupled matrix elements
of the interaction (\ref{eq:rint}), i.e.
\begin{widetext}
\begin{eqnarray}
A(ph,p'h')-(\e_{{n_p},{l_p}}-\e_{{n_h},{l_h}}) 
\de_{{l_p},{l_{p'}}} \de_{{l_h},{l_{h'}}}
\de_{{n_p},{n_{p'}}} \de_{{n_h},{n_{h'}}}
= 2 \; R(ph',hp') \; (-)^{l_p+l_{p'}}~~~~~~~~~~~
\nonumber \\
\times 
\frac { [(2l_p+1)(2l_h+1)(2l_{p'}+1)(2l_{h'}+1)]^\frac{1}{2} } {(2\la+1)}
\left( 
\begin{array}{ccc}
l_h & \la & l_p \\ 0 & 0 & 0 
\end{array}
\right)
\left( 
\begin{array}{ccc}
l_{h'} & \la & l_{p'} \\ 0 & 0 & 0 
\end{array}
\right) & &  \nonumber \\
= (-)^{\la} \; B(ph,p'h')~,~~~~~~~~~~~~~~~  \label{eq:mel}  
\end{eqnarray}
where
\beq
R(ph',hp')=\int r_1^2 dr_1 r_2^2 dr_2
{\cal R}_{{n_p},{l_p}}(r_1) 
{\cal R}_{{n_{h'}},{l_{h'}}}(r_2)
V(r_1,r_2; \la)
{\cal R}_{{n_h},{l_h}}(r_1) 
{\cal R}_{{n_{p'}},{l_{p'}}}(r_2) ~,  
\label{eq:rad}
\eeq
\end{widetext}
and where ${\cal R}_{{n_i},{l_i}}(r)$ 
is the radial part of 
single-particle wave functions. The radial contribution of the 
two-body interaction (\ref{eq:rint}) in multipole order $\la$
is given by 
\[
V(r_1,r_2; \la) =e^2\; \frac{ r_{<}^{\la} } { r_{>}^{{\la}+1} }
\;+\;\frac{dV_{xc}[\rho]}{d\rho}\;
\frac{\de(r_1-r_2)}{r_1^2}\; \frac{2\la+1}{4\pi}~,
\]
where $r_<=min(r_1,r_2)$ and $r_>=max(r_1,r_2)$.

The indices $n_i$ appearing in eqs. (\ref{eq:mel}) and (\ref{eq:rad})
denote the number of nodes for the corresponding single-particle
states. The orbital angular momenta of the particles and holes
participating in the excitations are denoted by $l_i$, the total
angular momentum of the excitation being $\la$, which in the
present calculation is set equal to 1 (dipole vibration).
The  3j symbols appearing in (\ref{eq:mel}) take proper care
of the angular momentum coupling, as well as of the parity
conservation conditions. The factor 2 accounts for the spin
degeneracy.

The RPA eigenvectors are
written as a linear combination of particle-hole excitations
in terms of the forwardsgoing and backwardsgoing amplitudes,
$X_n(ph)$ and $Y_n(ph)$ respectively, according to
\begin{eqnarray}
|n\rangle &=&
\sum_{ph} [ X_n(ph) |(ph^{-1})_{{\la}{\mu}} \rangle \nonumber \\
&& \hspace{0.7cm} -(-)^{\la+\mu} \; Y_n(ph) |(hp^{-1})_{{\la},-{\mu}} \rangle ]~,
\label{eq:eig}
\end{eqnarray}
where a singlet spin configuration is implied.
 
The dipole transition probabilities associated with the state
(\ref{eq:eig}) can be written as
\[
B(E1,0 \rightarrow n) = \frac{2}{3}\;|\langle n||{\cal M} (E1) ||0\rangle|^2~,
\]
where
\begin{widetext}
\beq
\langle n|| {\cal M} (E1) ||0\rangle =
\sum_{ph} \left[X_n^*(ph;1)+(-)^{\la} \; Y_n^*(ph;1) \right] \;
\langle p|| {\cal M} (E1) ||h\rangle~,
\label{eq:tprob2}
\eeq
\end{widetext}
are reduced matrix elements \cite{bm} of the dipole operator
${\cal M} (E1;\mu)= \sqrt{4\pi/3} \; e r {\cal Y}_{1\mu}(\hat{r})$.
 
The radial wave functions $R_{{n_i},{l_i}}$ are calculated by
diagonalizing the single-particle Hamiltonian $H_0$ in a basis
including ${\cal N}=25$ harmonic oscillator major shells.
(For the oscillator parameter of this basis, we have used
$\h \w =1.1\; eV$. The results do not, however, depend
on this choice.)
 
The associated unperturbed particle-hole excitations, examples of which
are displayed in Fig. 1, exhaust the Thomas-Reiche-Kuhn sum rule
$S(E1)=N\h^2e^2/2m$, that is
\[
\sum_{ph} (\e_{{n_p},{l_p}}-\e_{{n_h},{l_h}}) 
|\sqrt{2/3} \;\langle p|| {\cal M} (E1) ||h\rangle |^2 = S(E1)~,
\]
where $\e_{{n_i},{l_i}}$ are the single-particle energies. This result is
also valid for the correlated eigenstates (\ref{eq:eig}) and
associated eigenvalues $E_n$, since the RPA preserves
the energy-weighted sum rule (EWSR).
 
In Fig. 2 we display the oscillator strength functions versus excitation energy 
for the RPA dipole in the case of the neutral Na$_8$ and Na$_{20}$ clusters,
as well as in the case of the charged
Na$_9^+$ and Na$_{21}^+$ clusters. 
The unperturbed oscillator strength for the neutral clusters is also
displayed.

A conspicuous feature of the RPA results for the neutral clusters is the
sizeable amount of Landau damping. In fact,
the variances of these response functions
are $\sigma \approx 0.57\; eV$ in both cases implying a ratio
$\sigma/{\bar E} \approx 0.25$, ${\bar E}$ here being the
energy centroids of the
resonances. Nonetheless, the identification of the collective states
is quite unique. Indeed, in the case of Na$_8$, there is
a single state at $\approx 2.8\;eV$ exhausting $\approx 75\%$ of the
EWSR, while about the same strength is distributed among two
lines located at $2.6 \; eV$ and $2.9\;eV$ in the case of Na$_{20}$.
As befitted collective states, the associated wave functions  
exhibit several (about ten) forwardsgoing amplitudes
(X-components) that are larger than 0.1. Also, most of the
backwardsgoing amplitudes (Y-components) contribute constructively
to the transition amplitude (\ref{eq:tprob2}).
The wave functions associated with the rather weak states which
are strongly red shifted with respect to the collective states
are, on the other hand, dominated by a couple of components. 
%
%
We note that the results shown in the middle of Fig. 2 for Na$_{20}$
display an extent of fragmentation similar to the TDLDA calculation by
Ekardt (cf. Fig.6 of ref. \cite{ekar}).
 
The plasmon strength function is drastically modified in the case of
the charged Na$_9^+$ and Na$_{21}^+$ clusters, where a single peak
lying at $\approx$ 3.1 $eV$ (Na$_9^+$) and at $\approx$ 3.0 $eV$
(Na$_{21}^+$) exhausts $\approx$ 93\% and $\approx$ 83\% of the EWSR,
respectively. This result seems to be consistent with recent
experimental findings reported\cite{orsa} for the potassium clusters
K$_9^+$ and K$_{21}^+$.

The difference between positively charged and neutral clusters is due to
the difference in the associated average potentials (cf. Fig. 1).
Indeed, the potentials for the charged clusters are
deeper than the potentials for the corresponding neutral clusters
(observe that the potential for the neutral clusters 
exhibit an almost constant central depth at $\approx -5.4$ $eV$).       
For Na$_9^+$ the increase of the potential 
depth at the center amounts to $\approx$ 56\%, while for Na$_{21}^+$ 
the corresponding increase is $\approx$ 42\%.

\begin{figure}[t]
\centering\includegraphics[width=8cm]{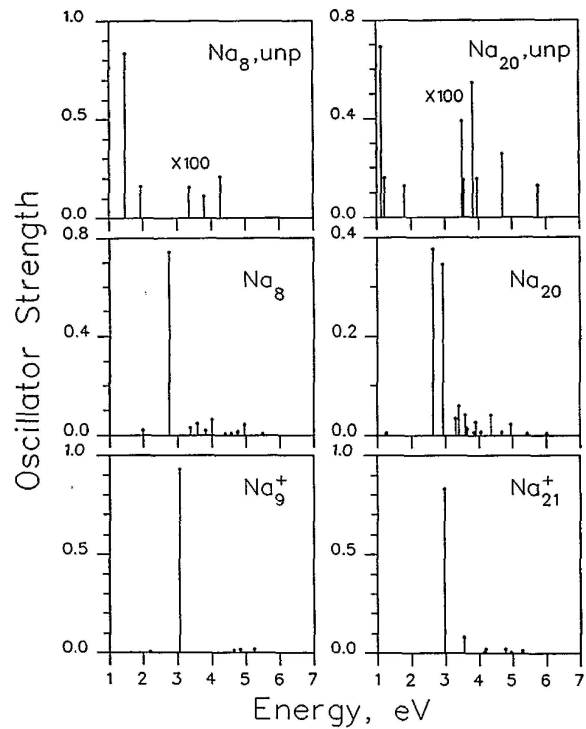}
\caption{
Oscillator strength function for the photoabsorption
of Na$_8$, Na$_{20}$, Na$_9^+$, and Na$_{21}^+$ clusters.
In the upper figures, the unperturbed oscillator strengths
for the neutral clusters is displayed.
Also, in the upper figures, the strength values above 3 $eV$ have been
multiplied by 100.
}
\end{figure}

It is interesting to compare the moments of the dipole strength
function obtained in the present microscopic RPA calculation 
for the case of neutral clusters with
those obtained semiclassically
in ref. \cite{brac}. The cubic energy-weighted sum
rules and the centroids ${\bar E}$ agree within $2-4\%$; the
static dipole polarizabilites $\alpha$ (given
by the negative energy-weighted sum rules) agree within
$4-10\%$. This offers support for the ability of the semiclassical 
method proposed in
ref. \cite{brac} to predict the average properties
of the mean-field response function.

\begin{figure}[t]
\centering\includegraphics[width=8cm]{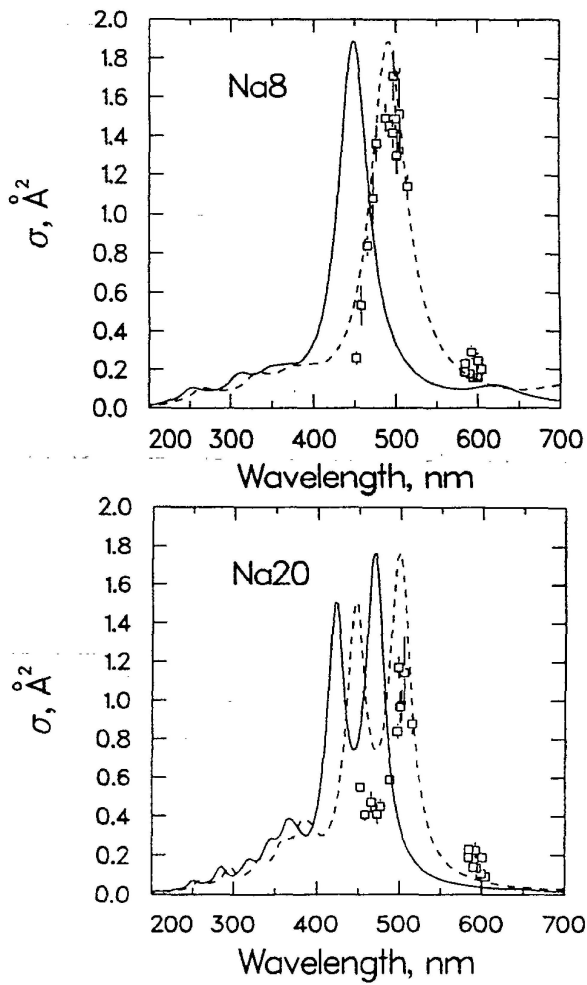}
\caption{
Photoabsorption cross sections per atom (for Na$_8$
and Na$_{20}$) resulting from
the folding of the RPA oscillator strengths with Lorentzian
shapes normalized to unity are shown (solid line).  The widths
of the Lorentzians are specified by $\G/\h\wo \approx 0.1$ and 0.06
for Na$_8$ and Na$_{20}$, respectively \cite{pb},
$\h\wo$ being the peak energy
of a given Lorentzian. The experimental data\cite{svmk} are
also shown. The dashed line is the solid line shifted so that
there is a maximum overlap with the data.
}
\end{figure}

In order to compare the present calculations with the
experimental data, one has to take into account the coupling
of the electronic dipole oscillations to thermal surface 
fluctuations of the whole cluster \cite{bt,pb}. 
This is done only for the case of neutral clusters
for which the most systematic body of information is available
at present\cite{foot}. 
In keeping with
the findings of ref. \cite{pb}, this can be approximately
simulated by folding the peaks of the RPA spectrum
with Lorentzian functions displaying a damping factor
$\G/\h\wo \approx 0.1$ and 0.06 for Na$_8$ and Na$_{20}$,
respectively. The results are shown in Fig. 3 in comparison with
the experimental data \cite{svmk}.
Aside from a shift of 
$\approx 10\%$, the overall structure
observed in these two cross sections is well reproduced. In particular
for Na$_{20}$, the dip seen in the experimental cross section
near 460 $nm$, as well as the weak bump in Na$_8$ near 590 $nm$,
seem to come out of the present RPA description.
 
The shift of the calculated dipole peaks with respect to the observed
values for the neutral clusters is consistent with the fact that the predicted
static dipole polarizabilities within the LDA model
\cite{eka2,brac,bek2} are lower than the experimental ones by 
$\approx 20\%$. This is because the model, when applied to the exchange
part of the Coulomb interaction between the electrons, provides
too much screening, and thus fails to produce for neutral clusters the expected
asymptotic $1/r$ Coulomb behavior of the mean field.  
As a consequence, the tail of the electron densities
(the so-called 'spill-out') is underestimated; since the latter is
known to be correlated to both the dipole polarizabilities \cite{bek2}
and the surface plasmons \cite{brac}, these quantities are also underestimated.
Indeed, a 'self-interaction correction' aimed at a better treatment
of the Coulomb exchange within the Kohn-Sham approach \cite{sb}
can improve the values for the dipole polarizabilities.
It remains an open question to which extent an improved treatment of the
Coulomb exchange will affect the positions of
the unperturbed particle-hole excitations, and thus
the detailed structures of the photoabsorption cross sections
obtained in the present investigation.
 
It can be concluded that direct decay of the giant dipole resonance
in small neutral sodium clusters plays an important role in the
fragmentation of the associated strength function, a process that seems
to be essentially absent in the case of the positively charged clusters.
The actual location and potential fragmentation of the dipole peak
in Na$_8$ and Na$_{20}$ is the result of a delicate balance
between shell structure and residual interaction, which
might be further unravelled by carrying photoabsorption measurements at
low temperatures. A detailed mapping
of the strongly red shifted dipole peaks and of their structure
could provide some crucial tests of 
the exchange (plus correlation) part of the effective
interaction between the electrons in metal clusters.
 
Discussions with W. de Heer, B.R. Mottelson and H. Nishioka are
gratefully acknowledged. We wish to thank W.D. Knight and collaborators
for providing us with the most recent experimental data prior to
publication. One of us (C.Y.) wishes to acknowledge financial support
during the course of this research 
from the INFN, sez. Milano, from a NATO fellowship through the Greek
Ministry of National Economy, and from the Joint Institute for Heavy
Ion Research, where the calculations were completed.


\begin{thebibliography}{99}
\bibitem{foot}
In any case, thermal fluctuations are expected to 
be less important for charged clusters due to the fact that the
electrons are more tightly bound, and thus the associated shell
corrections\cite{brac2} (Nilsson-Strutinsky model) 
are expected to be larger leading to a stiffer potential energy
surface in the space of cluster shapes, as compared to the neutral
case.
\end{thebibliography}
\end{document}